\documentclass[reprint, superscriptaddress, twocolumn, amsmath, amssymb, pra]{revtex4-2}

\usepackage{graphicx}
\usepackage{dcolumn} 
\usepackage{bm}
\usepackage[colorlinks=true, citecolor=magenta, linkcolor=magenta]{hyperref}

\usepackage{mathrsfs}
\usepackage{color}
\usepackage{soul}
\usepackage{braket}

\usepackage{lipsum}
\usepackage{xargs}
\usepackage[colorinlistoftodos,prependcaption,obeyDraft]{todonotes}
\usepackage{setspace}

\usepackage{float}
\usepackage{booktabs}

\definecolor{capri}{rgb}{0.0, 0.45, 0.73}
\definecolor{bluebell}{rgb}{0.64, 0.64, 0.82}

\newcommand{\Op}[1]{\ensuremath{\hat{#1}}}

\newcommand{\Avg}[1]{\ensuremath{\langle #1 \rangle}}

\renewcommand{\Re}{\mathrm{Re}}

\begin{document}


\title{Time-reversal Interferometry Using Cat States with Scalable Entangling Resources}

\author{Sebasti\'an C. Carrasco}
 \email{seba.carrasco.m@gmail.com}
\affiliation{
DEVCOM Army Research Laboratory, Adelphi, MD 20783
}%

\author{Michael H. Goerz}
\affiliation{
DEVCOM Army Research Laboratory, Adelphi, MD 20783
}%

\author{Zeyang Li}
\affiliation{
MIT-Harvard Center for Ultracold Atoms and Research Laboratory of Electronics,
Massachusetts Institute of Technology, Cambridge, MA 02139
}%
\affiliation{Department of Applied Physics, Stanford University, Stanford, CA}

\author{Simone Colombo}
\affiliation{
MIT-Harvard Center for Ultracold Atoms and Research Laboratory of Electronics,
Massachusetts Institute of Technology, Cambridge, MA 02139
}%
\affiliation{Department of Physics, University of Connecticut,
196A Auditorium Road, Unit 3046, Storrs, CT 06269-3046}

\author{Vladan Vuleti\'c}
\affiliation{
MIT-Harvard Center for Ultracold Atoms and Research Laboratory of Electronics,
Massachusetts Institute of Technology, Cambridge, MA 02139
}%

\author{Wolfgang P. Schleich}
\affiliation{Institute of Quantum Physics and Center for Integrated Quantum Science and Technology (IQST), Ulm University, Ulm, Germany}

\author{Vladimir S. Malinovsky}
\affiliation{
DEVCOM Army Research Laboratory, Adelphi, MD 20783
}%

\date{\today}

\begin{abstract}
We propose a novel method for generating Schr\"odinger-cat states---defined as equal superpositions of arbitrary coherent states---using a concise sequence of rapid twist-and-turn pulses. We demonstrate that the required shearing strength for the protocol, which scales linearly with time, decreases with increasing number of atoms ($N$) in proportion to $1/\sqrt{N}$. The resulting states exhibit optimal quantum Fisher information, making them ideal for surpassing the classical limit of phase sensitivity in quantum metrology applications. Notably, our protocol is compatible with a time-reversal strategy for quantum metrology, ensuring its practical viability. Furthermore, we demonstrate that the Heisenberg limit scaling remains intact even when reducing the twisting employed in tandem with the number of atoms, thereby mitigating realistic losses such as photon scattering.
\end{abstract}

\maketitle

\section{Introduction}

High-precision measurements are crucial in various research areas, including testing fundamental laws~\cite{safronova2018search, safronova2019search}, general relativity~\cite{SannerN2019, KennedyPRL2020}, quantum computing~\cite{XuerebPRL2023}, and searching for new physics~\cite{derevianko2014hunting}. Achieving higher measurement precision can impact diverse fields such as gravitational wave detection~\cite{kolkowitz2016gravitational}, timekeeping~\cite{GuinotM2005}, and geodesy~\cite{mehlstaubler2018atomic, grotti2018geodesy, takamoto2020test}. However, the discrete nature of light and matter limits measurement error to the standard quantum limit (SQL), which scales as $1/\sqrt{N}$ with the number of probes $N$. Quantum effects, such as entanglement, can surpass this limit, enabling quantum metrology to achieve the Heisenberg limit (HL) scaling, where measurement errors scale as $1/N$~\cite{degen2017quantum}. This allows for ultra-high-precision sensing with relatively few atoms by exploiting quantum properties.

The capability to attain HL scaling depends on quantum correlations in the generated multi-atomic states used as a probe. Non-correlated states, such as coherent spin states (CSS), can only achieve SQL scaling, while maximally entangled states, such as Greenberger-Horne-Zeilinger (GHZ) states, enable HL scaling. However, generating and manipulating GHZ states is challenging due to decoherence, dissipation, and atom losses~\cite{LeibfriedN2005, PezzeRMP2018, NagibPRL2025}.

Fortunately, other states can also achieve sub-SQL scaling~\cite{MalekiPRA2021, GaoNP2010, LeibfriedS2004}, such as Schr\"odinger-cat states~\cite{WinelandRMP2013}, which have higher immunity to decoherence~\cite{SandersPRA2014, HuangSR2015, HuangPRA2018, MalekiJOSAB2020}. In contrast to the GHZ state, a Schr\"odinger-cat state is a superposition of two CSSs that are not necessarily orthogonal. Thus, Schr\"odinger-cat-sates can be considered as a generalization of GHZ states.
Other states that can also attain sub-SQL scaling include spin-squeezed states~\cite{kitagawa1993squeezed, wineland1992spin, wineland1994squeezed, MaPR2011, CarrascoPRA2022, carrasco2023dicke}.
These states exhibit reduced variance in one measured quadrature, while increasing variance in an orthogonal measurement. Both the Schr\"odinger-cat-sates and spin squeezed states can be generated using pulse sequences of one-axis twisting (OAT)~\cite{kitagawa1993squeezed, MaPR2011} and rotations~\cite{Julia-diazPRA2012, SarkarPRA2018, KaubrueggerPRX2021, CarrascoPRA2022, HuangPRA2022, SongS2019}.  

In this work, we develop a novel scheme to generate Schr\"odinger-cat states using a pulse sequence of OAT and rotations with finite entangling resources. The time required  to generate these states favorably decreases  as $1/\sqrt{N}$, unlike the time-evolution under a pure OAT Hamiltonian, which also can generate Schr\"odinger-cat states in $N$ independent time~\cite{AgarwalPRA1997, SarkarPRA2018}. Additionally, we propose a time-reversal interferometric protocol in which the many-atom quantum system undergoes a pulse sequence, free evolution, and an inverse pulse sequence~\cite{SchulteQ2020, HostenS2016, LeibfriedS2004, LinnemannPRL2016, BurdS2019, colombo2021time, LiSCI2025, LiuPRL2025, ZaporskiN2025}. This sequence amplifies the phase acquired during free evolution, yielding Heisenberg-limited scaling. Unlike standard spin squeezing for Ramsey interferometry, our approach doesn't reduce variance to generate enhancement, making it immune to detection noise limitations~\cite{HostenS2016}. Our method is versatile and applicable to any OAT setup, using only engineered atom-atom interactions and standard rotation pulses. We focus on implementing this scheme in an optical-lattice atom-clock experiment, where a time-reversal interferometric protocol has been demonstrated~\cite{colombo2021time, LiSCI2025}. In this setup, photons incoherently scattered during the generation of the OAT Hamiltonian project sub-ensembles of atoms into ground and excited states, causing coherence loss and reducing the contributing atom number at the final measurement~\cite{LiPRXQ2022, colombo2021time}. Our analysis accounts for this loss effect, which diminishes measurement precision. By employing the proposed scheme, we demonstrate the generation of  Schr\"odinger-cat states and achieve HL scaling, which will enable ultra-high-precision sensing in various research areas.

The outline of our paper is as follows. In Sec. \ref{cat-states}, we characterize the Schr\"odinger-cat states in terms of the quantum Fisher information (QFI) and provide a proof of what Schr\"odinger-cat states are optimal for quantum metrology. In Sec. \ref{generation}, we prove which are the necessary controls to create Schr\"odinger-cat states and find pulse sequences to generate them. In Sec. \ref{protocol}, we present a time-reversal interferometric protocol capable of achieving HL scaling by creating and dismantling the Schrödinger-cat-states to amplify interferometric signals. A final discussion is in Sec. \ref{discussion}.

\section{Characterizing Schr\"odinger-cat states}
\label{cat-states}

We consider an ensemble of $N$ spin-$1/2$ particles, or equivalently, $N$ identical two-level systems. The collective spin operator $\vec{S} = [\Op{S}_x, \Op{S}_y, \Op{S}_z]$ is defined by its components $\Op{S}_i = \frac{1}{2} \sum_{n = 1}^N \Op{\sigma}_{i, n}$, where $i = x, y, z$ and $\Op{\sigma}_{i, n}$ is the Pauli matrix acting on the particle $n$. We define a CSS as a state where all individual spins are in the same state; consequently, the collective spin state is 
\begin{equation}
  \ket{\theta, \phi} = \bigotimes_{\ell = 1}^{N}\, \left(\cos \frac{\theta}{2} \ket{\uparrow}_\ell + e^{i \phi} \sin \frac{\theta}{2} \ket{\downarrow}_\ell\right) \, ,
\end{equation}
where $\theta$ and $\phi$ define the direction in spherical coordinates where the individual spins are oriented.

Since a Schr\"odinger-cat state is a superposition of two CSSs $\ket{\theta_1, \phi_1}$ and $\ket{\theta_2, \phi_2}$, its wave function can be represented as 
\begin{equation}
  \ket{\Psi_{\text{cat}}} = \frac{1}{C}\left(\ket{\theta_1, \phi_1} + \ket{\theta_2, \phi_2}\right) \,, 
\end{equation}
where $C^2=2\Re(1 + C_N)$ is the normalization factor, $C_N= \left(\mathcal{C}_1 \mathcal{C}_2 +  e^{i\Delta \phi} \mathcal{S}_1 \mathcal{S}_2 \right)^N$, $\Delta \phi=\phi_2-\phi_1$, $\mathcal{C}_i=\cos (\theta_i/2)$, 
$\mathcal{S}_i=\sin (\theta_i/2)$.

To quantify the potential of Schr\"odinger-cat states for quantum metrology, we evaluate the quantum Fisher information (QFI), which for pure states is defined by 
\begin{equation} \label{eq:QFI}
    \mathcal{F}_Q = 4 \Delta \Op{S}_z^2 \, ,
\end{equation}
with $\Delta \Op{\mathcal{O}}^2 = \Avg{\Op{\mathcal{O}}^2} - \Avg{\Op{\mathcal{O}}}^2$. The quantum Cram\'er-Rao bound (qCRB), $\Delta \varphi^2 = 1/\mathcal{F}_Q$, defines the minimum uncertainty that can be achieved for an estimation of $\varphi$ using a given collective spin state $\ket{\Psi_\text{cat}}$. The phase $\varphi = \omega \tau$ is the one acquired while the system evolves under the Hamiltonian $\hat H_0 = \omega \hat S_z$ for a time $\tau$, for instance, during the free evolution in a Ramsey interferometry procedure.

For an arbitrary Schr\"odinger-cat state, we obtain
\begin{equation} \label{Eq:4}
    \braket{\Op{S}_z} = \frac{N}{C^2}  \Re\left[\mathcal{C}_+ \mathcal{C}_- + C_{N-1} \left( 
 \mathcal{C}_1 \mathcal{C}_2 - e^{ i \Delta \phi} \mathcal{S}_1 \mathcal{S}_2 \right)
\right]
\end{equation}
and
\begin{multline} \label{Eq:5}
    \braket{\Op{S}_z^2} = \frac{N}{4} + \frac{N (N-1)}{4 C^2}
  \Re \Big[ \cos^{2} \theta_1 + \cos^{2} \theta_2  \\ 
+ 2 C_{N-2} \left( 
 \mathcal{C}_1 \mathcal{C}_2 - e^{ i \Delta \phi} \mathcal{S}_1 \mathcal{S}_2 \right)^2
  \Big] \,,
\end{multline}
where $\mathcal{C}_{\pm}=\cos \left[(\theta_1 \pm \theta_2)/2\right]$, $C_k= \left(\mathcal{C}_1 \mathcal{C}_2 +  e^{i\Delta \phi} \mathcal{S}_1 \mathcal{S}_2 \right)^k$. 

Given a great-circle angle $\theta$ between each CSS, the Schr\"odinger-cat state that is symmetric with respect to the $xy$ and $xz$ planes maximizes the QFI as defined by Eq.~\eqref{eq:QFI}. Under these assumptions, we have
\begin{equation} \label{eq:cat}
  \ket{\Psi_\text{cat}(\theta)} = \frac{\ket{\pi/2 - \theta/2, 0} + \ket{\pi/2 + \theta/2, 0}}{\sqrt{2+2\cos^N(\theta/2)}} \, .
\end{equation}
We prove that this state is a QFI maximum in the Appendix \ref{Appendix:A}.

When considering the state in Eq.~\eqref{eq:cat}, we obtain
\begin{equation}
    \mathcal{F}_Q = N + N (N-1) \frac{\sin^2(\theta/2)}{1 + \cos^N(\theta/2)} \,.
\end{equation}
Similar expressions for general Schr\"odinger-cat states can be found for the QFI in arbitrary directions using the complementary expressions that we write in the Appendix \ref{Appendix:B}.
If $\theta = 0$, we have a CSS and the minimum uncertainty possible is $\Delta \varphi = N^{-1/2}$, corresponding to the SQL. If $\theta \neq 0$, we can take the limit $N \gg 1$, to obtain
\begin{equation} \label{Large-cat}
    \Delta \varphi^2 = \frac{1}{\mathcal{F}_Q} \approx \frac{1}{N^2 \sin^2 (\theta/2)} \, ,
\end{equation}
which exhibits HL scaling, even for an arbitrarily small value of $\theta$, when the overlap between CSSs, $\braket{\pi/2 - \theta/2, 0| \pi/2 + \theta/2, 0}$, is very large. Note that $\theta = \pi$ gives exactly the HL, as expected from a GHZ state. One can understand this from pure geometrical considerations; in the limit $N \gg 1$, the average projection into the z-axis for the CSS in the northern hemisphere is $(N/2) \sin(\theta/2)$. Similarly, the average projection into the z-axis for the CSS in the southern hemisphere is $-(N/2) \sin(\theta/2)$. Thus, the total standard deviation for the cat state is $N^2 \sin^2(\theta/2)/4$, and consequently the QFI $\mathcal{F}_Q = N^2 \sin^2(\theta/2)$, as we found in Eq.~(\ref{Large-cat}).

\section{Creating Schr\"odinger-cat states}
\label{generation}

\begin{figure}
    \centering
    \includegraphics{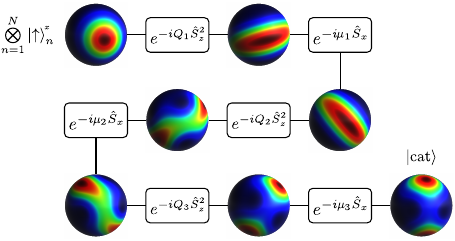}
    \caption{%
      \label{fig:sequence}
      Sequence example where $n=6$ pulses are used to create a Schr\"odinger-cat state with $\theta = \pi/2$ in $N = 20$ atoms. The infidelity respect to the target is $\epsilon = 0.02$.
    }
\end{figure}

We propose that alternating OAT applications with collective rotations is sufficient to drive a CSS to an optimal Schr\"odinger-cat state. To prove this affirmation, we introduce the Dicke basis $\ket{S, m}$, with $S = N/2$ and $m=-S, -S+1, \ldots, S-1, S$, corresponding to the  eigenfunctions of $\hat S_z$, namely, $\hat S_z \ket{S, m} = m \ket{S, m}$. 

We note that Schr\"odinger-cat states are symmetric on this basis, thus $\ket{\Psi_\text{cat}(\theta)} = \sum_{m=-S}^S a_m \ket{S, m}$ with $a_m = a_{-m}$. Thus, it is convenient to describe the dynamics to define the subspace of symmetrized Dicke states,
\begin{equation*} \label{} 
  \ket{S, m}_s =  
  \begin{cases}
    \frac{1}{\sqrt{2}} \left(\ket{S, m} + \ket{S, -m}\right) & \text{if $m \neq 0$} \,,\\
    \ket{S, 0} & \text{if $m = 0$}
  \end{cases} \, ,
\end{equation*}
On this basis,
\begin{equation*} \label{}
  \hat S_z^2 \ket{S, m}_s = m^2 \ket{S, m}_s \,,
\end{equation*}
and
\begin{multline*} \label{} 
  \hat S_x \ket{S, m}_s = \frac{1}{2} \sqrt{S(S+1) - m(m+1)} \ket{S, m+1}_s \\ + \frac{1}{2} \sqrt{S(S+1) - m(m-1)} \ket{S, m-1}_s \,,
\end{multline*}
which shows that these operators preserve $a_m = a_{-m}$ symmetry and time-evolution under them remains in the symmetrized subspace.

Now we turn to controllability. Let us consider $\Op H_0 = \Op S_z^2$ to be the drift Hamiltonian and $\Op H_1 = \Op S_x$ to be the control Hamiltonian. We can note three things: (i) $H_1$ couples the state $\ket{S, m}_s$ with $\ket{S, m+1}_s$ and $\ket{S, m-1}_s$. Consequently, all nearest neighbor states are coupled, and all states are accessible. (ii) all transitions are not degenerated, as the transition frequencies are $\omega_{m, m-1} = \bra{S, m}_s\hat S_z^2 \ket{S, m}_s - \bra{S, m-1}_s\hat S_z^2 \ket{S, m-1}_s = 2m - 1$, and are different for all values of $m$. (iii) all relevant (nearest neighbor) fractions of transition frequencies $\omega_{m, m-1} / \omega_{m', m'-1}$ are rational numbers. According to Refs. \cite{DongIETCTA2010, TuriniciCP2001, TuriniciJPAMG2003}, these three conditions are sufficient for pure state controllability. Perhaps the key point is that the $H_0 = \Op S_z^2$ nonlinearity removes the degeneracy and enables control.

Summing up, OAT and rotations around the x-axis are sufficient to create any state in this subspace. In particular, one can start from a CSS oriented to the x-axis $\ket{\pi/2, 0}$ (which also belongs to the symmetric subspace) and create an Schr\"odinger-cat state. Note that these controls are available in the experimental setup in which we are focusing on our approach.

We propose a sequence of $2 n$ alternating OAT and $S_x$ pulses that allow the creation Schr\"odinger-cat states from a CSS oriented to the $\hat x$ direction $\ket{\pi/2, 0}$. Figure \ref{fig:sequence}, illustrates the proposed pulse sequence with the state distribution over the generalized Bloch sphere for an example of $N=20$ atoms. Intuitively, OAT rotates the distribution on the generalized Bloch sphere with magnitude proportional to $z$. Thus, the northern and southern hemispheres rotate in opposite directions, with the magnitude being greatest at the poles. The first squeezing pulse creates a squeezed distribution, but it can break it into a two-peaked distribution upon reorienting the distribution appropriately (see Fig. \ref{fig:sequence}). Finally, by repeating this process appropriately, one can generate a target Schr\"odinger-cat state.

To find the pulse sequence that generates the objective Schr\"odinger-cat state, we propagate the initial CSS given values of $Q_k$ and $\mu_k$
\begin{equation}
\begin{split}
  \ket{\Psi_{\text{guess}}}
  &=
  \exp \left(- i \mu_{n} \Op{S}_x \right)
  \exp \left(- i Q_{n} \Op{S}_z^2 \right)\,
  \cdots
  \\
  &
  \exp \left(- i \mu_1 \Op{S}_x \right)
  \exp \left(- i Q_1 \Op{S}_z^2 \right)
  \ket{\pi/2, 0}\,,
\end{split}
\end{equation}
where $Q_k=\chi t_k$ is the shearing strength of $k$'th OAT pulse and $\mu_k$ is the rotation angle produced by the $k$'th $S_x$ pulse. Next, we estimate numerically the gradient of the infidelity $\epsilon = 1 - |\braket{\Psi_{\text{cat}}(\theta)| \Psi_{\text{guess}}}|^2$ respect to $(Q_k, \varphi_k)$ using finite differences to feed a gradient-based optimization method~\cite{ZhuATMS97,ByrdSJSC1995}. In this way, we iteratively find the parameters $(Q_k, \mu_k)$ that minimize $\epsilon$.  For instance, the sequence in Figure \ref{fig:sequence} creates a Schr\"odinger-cat-sate with $\theta = \pi/2$ and infidelity $\epsilon \approx 0.02$.

\section{Interferometric protocol}
\label{protocol}
\begin{figure}
    \centering
    \includegraphics{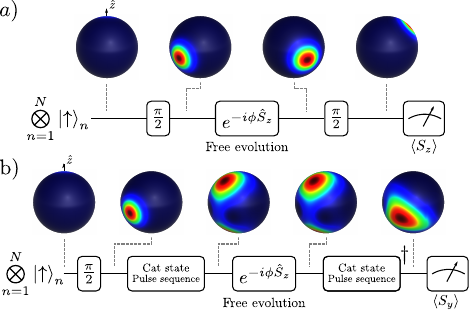}
    \caption{%
      \label{fig:protocol}
      (a) Schematic of the standard Ramsey interferometry scheme using a CSS.
      (b) Schematic of the modified Ramsey interferometry scheme using a Schr\"odinger-cat-sate.
    }
\end{figure}

In general, an interferometric protocol has three stages. The first is probe-state preparation, the second is free evolution, in which the system evolves under a free Hamiltonian, and the third is detection. An example of this is the Ramsey interferometry illustrated over the generalized Bloch sphere in Fig. \ref{fig:protocol}(a). In this protocol, the first stage is to prepare a CSS at the equator using a $\pi/2$ pulse. In the free evolution, the system evolves under the Hamiltonian $\hat H = \omega \Op S_z$ for a time $\tau = \varphi/\omega$. At last, in the detection stage, a $\pi/2$ pulse maps the phase, or the rotation angle, into the projection over the z-axis. Note that this pulse is the inverse of the first $\pi/2$ pulse, exhibiting a time-reversal quality. Finally, $\braket{\Op{S}_z}$ is measured and the result allows the estimation of $\varphi$ with SQL uncertainty.

In Fig. \ref{fig:protocol}(b), we illustrate the interferometric protocol that we propose. The state preparation consists of the Schrödinger-cat-state creation, and free evolution is under $\Op H = \omega \Op S_z$. However, we must determine appropriate detection, as simply applying a $\pi/2$ pulse before readout won't yield any population difference due to the probe state's symmetry. The approach we will consider here is time reversal. Consequently, we propose to apply the inverse of the pulse sequence that creates the Schr\"odinger-cat state after free-evolution, in analogy to the standard Ramsey interferometry. In this case, the last stage maps the phase to a variation of the expected value of $\Op S_y$ (which can be mapped to a population difference by a $\pi/2$ pulse). In the absence of an accumulated phase, the protocol returns the system to its initial state. Otherwise, it provides a state close to the starting CSS but with the displacement proportional to the accumulated phase.

The minimal resolvable phase of the protocol $\Delta \varphi$ can be estimated from $\Delta \Op S_y$ at the end of the protocol, namely,
\begin{equation}
  \label{eq:sensitivity}
  \Delta \varphi = \left| \frac{\Delta \Op{S}_y}{\partial \Avg{\Op{S}_y} / \partial \varphi} \right| \, .
\end{equation}
Thus the metrological gain respect to the SQL is given by
\begin{equation}
  \mathcal{G} = \frac{(\Delta \varphi)^2_\text{SQL}}{(\Delta \varphi)^2} = \frac{1}{N (\Delta \varphi)^2} \, .
\end{equation}
In this expression, HL scaling corresponds to an exponent $b=1$ in $\mathcal{G} = a N^{-b}$. As in previous time-reversal implementations~\cite{colombo2021time}, since the resulting state is close to being a CSS, metrological gain comes from a displacement surpassing the one from Ramsey interferometry. Thus, $\Avg{\Op S_y} = \mathcal{M} \varphi S$, with an amplification factor $\mathcal{M}$ surpassing unity.

In an optical lattice clock atom experiment~\cite{colombo2021time}, collective entanglement generation relies on the interaction of the atomic ensemble with a light cavity field. Cavity feedback squeezing~\cite{leroux2010implementation,LiPRXQ2022}  is a deterministic technique to generate an effective OAT Hamiltonian $\chi \Op{S}_z^2$. The spin quantum noise tunes the atom-cavity resonance such that the intracavity light intensity is proportional to $S_z$. However, any information contained in the light field results in the non-unitary evolution of the atomic system~\cite{liu2011spin}. In particular, a photon scattered in free space projects one atom into either spin-up or spin-down, thereby reducing spin coherence and the length of the spin vector. However, as it has been recently demonstrated~\cite{colombo2021time,LiPRXQ2022} and modeled~\cite{liu2011spin,ZhangPRA2015} that
decoherence can be minimized to reach a near-unitary evolution by tuning the entangling-light frequency. Finally,  by switching the sign of the detuning between the atom-cavity resonance and the light, the sign of $\chi$ can be reversed~\cite{LiPRXQ2022,liu2011spin}.

As in Ref. \cite{colombo2021time}, we parametrize the spin vector's length as $C_\text{sc} = \exp ( - \gamma \tilde Q)$, where $\tilde Q = \sqrt{N} \, Q$ is the normalized shearing strength and $\gamma$ is the scaling parameter which we choose to be equal to $0.36$ addressing the experimental conditions reported in~\cite{colombo2021time}. Consequently, larger values of $\tilde Q$ in the pulse sequence that creates Schr\"odinger-cat-sates result in a lower signal amplitude $\braket{S_y}$ at the measurement. To minimize this effect in the metrological gain scaling, we fix $\tilde Q$ during the optimization of the pulse sequence. As a result, the contrast loss becomes a proportionality factor and does not affect the scaling. Besides, it ensures a time scaling proportional to $1/\sqrt{N}$. However, this could reduce the fidelity of the prepared state.

\begin{figure}
    \centering
    \includegraphics{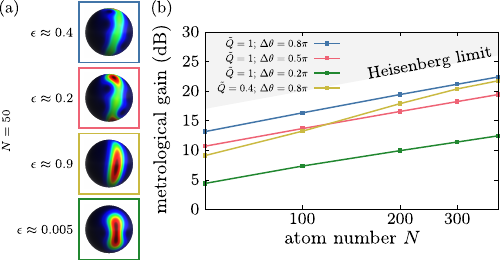}
    \caption{%
      \label{fig:scaling}
      Schr\"odinger-cat states metrological gain with fixed normalized shearing strength.
      (a) Examples of Schr\"odinger-cat states generated by pulse sequence with $n=3$ OAT and $\hat S_x$ pulses, with fixed total shearing strength $\tilde Q = \sqrt{N} \chi t$, for $N = 50$.
      (b) Metrological gain when targeting Schr\"odinger-cat states with different angles $\Delta \phi$ with fixed shearing strength $\tilde Q$ using $n=3$ OAT and $\hat S_x$ pulses.
    }
\end{figure}
%

Figure \ref{fig:scaling}(b) shows the achieved metrological gain for different values of $\theta$ and $\tilde Q$. The scaling exponent for the metrological gain with a normalized shearing strength $\tilde Q = 1$  ranges between $0.9$ (for $\theta = 0.2 \pi$) and $1.2$ (for $\theta = 0.8 \pi$), therefore, surrounding HL scaling. Note that it is possible to have faster scaling than the HL for a range of $N$ values. What is not possible is for the meteorological gain to exceed the HL (grey zone in the plot). When reducing the normalized shearing strength $\tilde Q$ to 0.4, we observe an important reduction in the metrological gain for low values of $N$. Yet, it approaches the result with $\tilde Q = 1$ when increasing $N$. When considering the loss due to photon scattering, $\tilde Q = 1$ corresponds to a loss of 6.4 dB, while $\tilde Q = 0.4$ corresponds to a loss of 2.5 dB. Consequently, the case with $\tilde Q = 0.4$ turns out to be better for these parameters. 

Figure \ref{fig:scaling}(a) illustrates the prepared states in the generalized Bloch sphere for $N=50$, for the same parameters of Figure \ref{fig:scaling}(b). As we can observe, the pulses reach the target state with lower infidelity for low values of $\theta$ and higher values of $\tilde Q$. Nevertheless, as the pulse sequence attempts to create Schr\"odinger-cat states, it increases $\Delta S_z$ in the process and therefore the Fisher information $\mathcal{F}_Q$, explaining why the metrological gain is still high with moderate infidelities.   

\section{Discussion and Conclusion}
\label{discussion}

We have demonstrated that the appropriate OAT pulses and rotations can generate any symmetric state in the Dicke basis, including  symmetric Schr\"odinger-cat states with respect to the equatorial plane of the generalized Bloch sphere. These states have attracted some attention because of their squeezing, planar squeezing~\cite{birrittella2021optimal}, and their potential in the framework of quantum computing~\cite{MirrahimiNJP2014}. Moreover, Schr\"odinger-cat states are also useful for ultra-high-sensitivity sensing and metrology under the appropriate protocol because of their HL scaling Fisher information~\cite{MalekiPRA2021, SarkarPRA2018}.

We developed a time-reversal interferometric protocol and analyzed its performance, focusing on potential implementation in an optical lattice clock atom experiment. The protocol employs a short pulse sequence of alternating OAT and rotation pulses to create Schrödinger-cat states, and then uses the opposite pulse sequence after free evolution, thus creating a protocol in which any accumulated phase breaks time-reversal symmetry. Our results show HL scaling even when considering contrast loss due to the same light-mediated interactions that create the squeezing, i.e., limiting the shearing strength of the OAT applications. The protocol uses the identical elements as existing OAT implementations and provides a time scaling proportional to $1/\sqrt{N}$, which is certainly favorable. Thus, it does not require major modification to an existing practical setup.

\section*{Acknowledgments}
This research was supported by the Army Research Laboratory under Cooperative Agreement Numbers W911NF-24-2-0044 (SC) and W911NF-23-2-0128 (MG).
VSM is grateful for support by a Laboratory University Collaboration Initiative (LUCI) grant from OUSD.

\appendix

\section{Schr\"odinger-cat states maximize QFI} \label{Appendix:A}

To show that the state in Eq.~\eqref{eq:cat} maximizes QFI, we compute it how changes when rotating the state using the unitary $\Op R = e^{-i \alpha \Op S_x - \beta \Op S_y}$. Thus, the QFI after rotation becomes $\mathcal{F}_Q = 4 \Avg{\Op R \, \Op S_z^2 \, \Op R^\dagger} - 4 \Avg{\Op R \, \Op S_z \, \Op  R^\dagger}^2$. Expanding up to quadratic order assuming $\alpha, \beta \ll 1$, we obtain
\begin{align}
    \Avg{\Op R \, \Op S_z \, \Op  R^\dagger} &\approx \Avg{\Op S_z} + \beta \Avg{\Op S_x} - \alpha \Avg{\Op S_z} - \frac{\alpha^2 + \beta^2}{2} \Avg{\Op S_z} \nonumber \\
     &\approx 0 \, ,
\end{align}
due to the symmetry of the state. Similarly,
\begin{multline}
    \Avg{R \Op S_z^2 R^\dagger} \approx (1 - \alpha^2 - \beta^2) \Avg{\Op S_z^2} + \alpha^2 \Avg{\Op S_y^2} + \beta^2 \Avg{\Op S_x^2} \\ \hspace{0.5 cm} - \alpha \Avg{\{\Op S_y, \Op S_z\}} + \beta \Avg{\{\Op S_x, \Op S_z\}} - \alpha \beta \Avg{\{\Op S_x, \Op S_y\}} \, .
\end{multline}
This reduces to
\begin{equation}
    \Avg{R \Op S_z^2 R^\dagger} \approx (1 - \alpha^2 - \beta^2) \Avg{\Op S_z^2} + \alpha^2 \Avg{\Op S_y^2} + \beta^2 \Avg{\Op S_x^2} \, .
\end{equation}
Given the symmetry of the state, for any pair of axes $u$ and $v$, we have $\Avg{(\Op S_u + \Op S_v)^2} = \Avg{(\Op S_u - \Op S_v)^2}$. This implies that
\begin{equation}
    \Avg{\{\Op S_u, \Op S_v\}} = \Avg{(\Op S_u + \Op S_v)^2} - \Avg{(\Op S_u - \Op S_v)^2} = 0 \, ,
\end{equation}
Consequently, we can approximate up to second order
\begin{equation}
    \mathcal{F}_Q = 4 (1 - \alpha^2 - \beta^2) \Avg{\Op S_z^2} + 4\alpha^2 \Avg{\Op S_y^2} + 4\beta^2 \Avg{\Op S_x^2}
\end{equation}
We conclude that corrections to the QFI are quadratic and with negative concavity (since $\Avg{\Op S_z^2} > \Avg{\Op S_{x, y}^2}$). Consequently, symmetric Schr\"odinger-cat states maximize QFI.

\section{Expectation values for arbitrary Schr\"odinger-cat states} \label{Appendix:B}

For an arbitrary Schr\"odinger-cat state, we obtain the following expressions that complement Eq.~\eqref{Eq:4} and \eqref{Eq:5},
\begin{multline}
   \braket{\Op{S}_x} 
=\frac{N}{2C^2}  \Re \left[\sin \theta_1 \cos \phi_1 + \sin \theta_2 \cos \phi_2 \right. \\ \left. + 2 C_{N-1} \left(e^{i \phi_1} 
\mathcal{S}_1 \mathcal{C}_2 + e^{i \phi_2} \mathcal{C}_1 \mathcal{S}_2 \right)
\right]
\,, 
\end{multline}
\begin{multline}
   \braket{\Op{S}_y} 
=\frac{N}{2C^2}  \Re \left[\sin \theta_1 \cos \phi_1 + \sin \theta_2 \cos \phi_2 \right. \\ \left. - 2i C_{N-1} \left(e^{i \phi_1} 
\mathcal{S}_1 \mathcal{C}_2 + e^{i \phi_2} \mathcal{C}_1 \mathcal{S}_2 \right)
\right]
\,, 
\end{multline}
\begin{multline}
   \braket{\Op{S}_x^2} 
= \frac{N}{4} + \frac{N(N-1)}{4C^2}  \Re \left[\sin^2 \theta_1 \cos^2 \phi_1 + \sin^2 \theta_2 \cos^2 \phi_2 \right. \\ \left. + 2 C_{N-1} \left(e^{-i \phi_1} 
\mathcal{S}_1 \mathcal{C}_2 + e^{i \phi_2} \mathcal{C}_1 \mathcal{S}_2 \right)
\right]
\,, 
\end{multline}
and
\begin{multline}
   \braket{\Op{S}_y^2} 
= \frac{N}{4} + \frac{N(N-1)}{4C^2}  \Re \left[\sin^2 \theta_1 \sin^2 \phi_1 + \sin^2 \theta_2 \sin^2 \phi_2 \right. \\ \left. + 2 C_{N-1} \left(e^{-i \phi_1} 
\mathcal{S}_1 \mathcal{C}_2 + e^{i \phi_2} \mathcal{C}_1 \mathcal{S}_2 \right)
\right]
\,.
\end{multline}

\bibliographystyle{apsrev4-2.bst}
\bibliography{apssamp}

\end{document}